\documentclass[twocolumn,showpacs,floatfix,superscriptaddress,amsmath,amssymb,prb]{revtex4}
\usepackage{mathrsfs}
\usepackage{txfonts}
\usepackage{amssymb}
\usepackage{graphicx}
\usepackage{hyperref}
\usepackage{ulem}
 \usepackage{overpic}
 \usepackage{psfrag}
\usepackage{tabularx}
\usepackage{array}
\usepackage{placeins}

\begin{document}
\title{Topological quantum phase transition in bond-alternating spin-1/2 Heisenberg chains}

 \author{Hai Tao Wang}
\affiliation{Centre for Modern Physics and Department of Physics,
Chongqing University, Chongqing 400044, The People's Republic of
China}

 \author{Bo Li}
\affiliation{Centre for Modern Physics and
Department of Physics, Chongqing University, Chongqing 400044, The
People's Republic of China}

 \author{Sam Young Cho}
 \email{sycho@cqu.edu.cn}
\affiliation{Centre for Modern Physics and Department of Physics,
Chongqing University, Chongqing 400044, The People's Republic of
China}

\begin{abstract}
 We investigate string correlations in an
 infinite-size spin-1/2 bond-alternating Heisenberg chain.
 By employing the infinite matrix product state representation
 with the infinite time evolving block decimation method,
 a finite string correlation for extremely large lattice distances
 is  {\it directly} observed, contrast to
 an extrapolated extreme value for finite size chains.
 We find that a topological quantum phase transition
 occurs between two different phases separated
 and characterized by two different long-range string orders in the space of
 bond-alternating interactions.
 Also, the critical exponent $\beta$ from the long-range string
 orders is obtained as $\beta=1/12$
 and the central charge at the critical point is obtained as $c \simeq 1$,
 which shows that
 the topological phase transition belongs to the Gussian universality class.
 In addition, it is shown that, for the topological quantum phase transition,
 the phase boundary can be captured by the singular behavior of the von
 Neumann entropy and the pinch point of the fidelity per site.
\end{abstract}
% 75.10.Jm Quantized spin models
% 75.40.Mg Numerical simulation studies
% 75.30.Kz Magnetic phase boundaries (including magnetic transitions, meta-magnetism, etc.)
% $---------------------$

\maketitle

\section{Introduction}

 Landau's symmetry breaking theory~\cite{Landau2} (LSBT) for
 understanding the continuous phase transitions is one of the
 cornerstone theories in quantum many-body systems. Most phases and phase
 transitions could be understood by a local order parameter
 characterizing a symmetry breaking ~\cite{Anderson,Coleman}.
 Recently,
 a new type of phase transitions, for instance,
 appearing in the discovery of the quantum Hall
 effect~\cite{Klit}, could not be understood well by local order parameters.
 Such quantum phase transitions seem to occur between two different phases
 without any explicit symmetry breaking.
 A non-local order, i.e., a so-called topological order~\cite{Wen5},
 rather than local orders, enables us to
 characterize the two phases.
 The quantum phase transition is called
 topological quantum phase
 transitions (TQPTs)
 ~\cite{Wen1,Kitaev,Trebst,Hamma,Cast,Feng,Chen,Yu,JVidal2}.
 Topological orders have been intensively studied in various systems
 such as fractional quantum Hall
 states~\cite{Tsui,Laughlin},  Haldane phase~\cite{Haldane},
% topological insulators~\cite{Kane1,Kane2,Bern,Fu,Qi},
 chiral spin liquids~\cite{Kalmeyer,Wen6}, $Z_2$ spin
 liquids~\cite{Read,Wen7,Moessner}, and so on.

 In spin lattice systems, especially,
 a string order as a prototype example of non-local long-range order
 was introduced
 by Nijs and Rommelse~\cite{Nijs}, and Tasaki~\cite{Tasaki} and
 characterizes the Haldane phase.
 Since then, non-local string orders have been extensively applied to study
 quantum phase transitions in various spin systems such as anisotropic
 spin-$1$ Heisenberg chains~\cite{Alcaraz}, frustrated
 antiferromagnetic Heisenberg spin chains~\cite{Kolezhuk},
 spin-$1/2$~\cite{Hida,Hung1} and spin-$1$~\cite{Yamamoto}
 alternating Heisenberg chains, XXZ spin-$1$ chain~\cite{Su}, spin
 ladders~\cite{stringladder,Shelton}, spin tubes~\cite{Garlea}, and Kitaev
 spin model~\cite{Feng,Chen}, and so on.
 Also, a TQPT has been shown between two phases respectively characterized
 by different non-local string (topological) order  parameters
 in Kitaev spin models \cite{Feng,Chen}.

\begin{figure}
\includegraphics [width=0.25\textwidth]{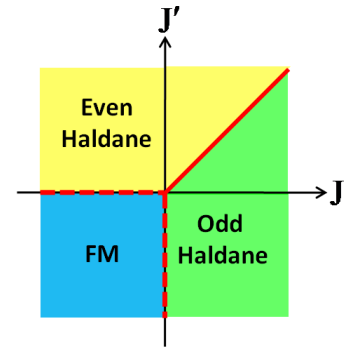}
\caption{ (Color online) Ground state phase diagram for the
 spin-$1/2$ bond-alternating Heisenberg chain in the plane of the bond alternations $J$ and
 $J'$.
 The even- and odd-Haldane phases are denoted by
 yellow and green, respectively. The ferromagnetic phase
 occurs in the blue region.
 Note that the red solid line between the two Haldane phases
 denotes a continuous phase
 transition corresponding to the topological phase transition,
 while the red dashed lines indicate a discontinuous
 phase transition.
 }
 \label{fig1}
\end{figure}

 In this paper, we study a TQPT in an
 infinite spin-$1/2$ bond-alternating Heisenberg chain
 by introducing
 two non-local string orders.
 To compute the two long-range string orders,
 we employ the infinite matrix product state (iMPS)
 representation~\cite{Vidal1,Vidal2}
 with the infinite time evolving block decimation (iTEBD) method
 developed by Vidal~\cite{Vidal2}.
 Compared to finite-size lattice systems where a long-range order
 determined by an extrapolated value of correlations,
 the iMPS approach allows us to calculate
 non-local long-range orders ~\cite{Su} directly.
 The string correlations and the dimer correlations directly
 calculated from the iMPS groundstate
 show clearly that there are two topologically ordered phases characterized
 by two long-range string orders.
 From the topological characterization,
 we find that there are three phases including
 the ferromagnetic phase in the spin-$1/2$ bond-alternating Heisenberg chain
 (See Fig. \ref{fig1}).
 Furthermore, the central charge $c\simeq 1$at the phase boundary between
 the two topologically ordered phases and the critical exponent of the string orders
 $\beta=1/12$ show
 that the TQPT belongs to Gussian-type phase transition.
 We discuss that
 the von Neumann entropy and the fidelity per lattice site (FLS)
 can capture the TQPT.

%%%%%%%%%%%%%%%%%%%%%%%%%%%%%%%%%%%%%%%%%%%%%%%%%%%%%%%
%%%%%%%%%%%%%%%%%%%%%%%%%%%%%%%%%%%%%%%%%%%%%%%%%%%%%%%

%
 This paper is organized as follows. In Sec.II, we introduce the
 spin-$1/2$ alternating Heisenberg chain model and discuss
 its groundstate energy.
 Section III devotes to discuss the topological phase transition
 by means of the comparison between
 the odd-/even-string correlations and the dimer correlations.
 In Sec. IV,
 the phase transitions are discussed from
 the von Neumann entropy. The TQPT is classified based on the central charge via the
 finite-entanglement scaling.
 The FLS is discussed to detect the TQPT in Sec. V.
 Finally, our conclusion is given in Sec. VI.

\section{spin-$1/2$ alternating Heisenberg chain and groundstate energy}
 The spin-$1/2$ alternating Heisenberg chains are described by the
 Hamiltonian~\cite{Black,Hida,Barnes}
\begin{equation}
H= \sum_{i} (J\vec{S}_{2i-1}\cdot\vec{S}_{2i}+
J'\vec{S}_{2i}\cdot\vec{S}_{2i+1}),
\label{Hamt}
\end{equation}
 where $\vec{S}=(S^x, S^y, S^z)$ are the spin-1/2 operators. $J$ and
 $J'$ are two alternative nearest-neighbor exchange couplings. In
 order to cover the whole range of the parameter $J$ and $J'$, we
 take $J=\cos\theta,J'=\sin\theta$. For $\theta = \pi/4$, this model
 is reduced to the conventional antiferromagnetic (AF) isotropic
 Heisenberg chain with being gapless, while for $\theta = 0$ or
 $\pi/2$, the system will be characterized by decoupled singlets.
 When $\theta$ approaches
 the limit $\theta=-\pi/2$ from $\theta=0$, Eq.~(\ref{Hamt}) can be
 regarded effectively as a $S=1$ AF isotropic Heisenberg chain, which
 can be characterized by a gapful Haldane phase.

%\section{Groundstate energy}
%
%%%%%%%%%%%%%%%%%%%%%%%%%%%%%%%%%%%%%%%%%%%%%%%%%%%%%%%%%
\begin{figure}
\includegraphics [width=0.45\textwidth]{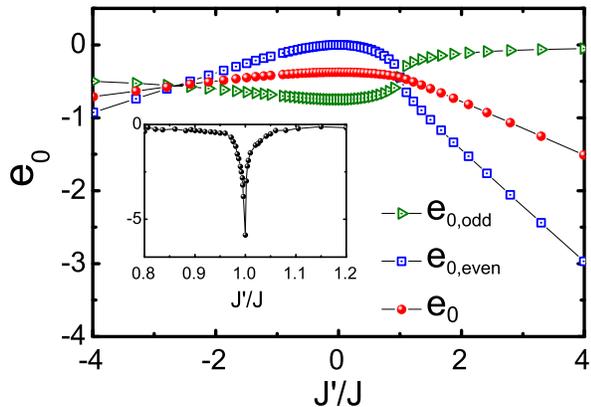}
\caption{ (Color online)
 Ground state energy per site $e_0$ and the odd-/even-bond
 energies $e_{0,even/odd}$ as a
 function of $J'$ with $J=1$.
 In the inset, the second order derivative of energy per site is
 plotted around the critical point $J'=J=1$. } \label{fig2}
\end{figure}
%%%%%%%%%%%%%%%%%%%%%%%%%%%%%%%%%%%%%%%%%%%%%%%%%%%%%%%%%
%
 From our iMPS groundstate in the Hamiltonian of Eq. (\ref{Hamt}),
 one can calculate the groundstate energy.
 In Fig~\ref{fig2}, the ground-state energy per site $e_0$, as an
 average value of the energies for odd bond $e_{0,odd}$ and even bond
 $e_{0,even}$, is plotted as a function of the alternation rate $J'/J$.
 We set the antiferromagnetic coupling $J=1$ and change
 $J'$. In the inset, the second-order derivative of the groundstate energy
 exhibits a singular behavior as $J'/J$ approaches $1$,
 which implies that the system undergoes a continuous
 (a second-order) phase transition across the critical point $J'/J=1$.

%%%%%%%%%%%%%%%%%%%%%%%%%%%%%%%%%%%%%%%%%%%%%%%%%%%%%%%%%
\begin{table}[b]
\caption{\label{tab:energy}
 Ground state energy per site $E_0$}
\begin{ruledtabular}
\begin{tabular}{clc}
Dimerization $\delta$ & DMRG $E_0$\footnote{Ref.~\onlinecite{Kumar}.} & iTEBD $E_0$($\chi=56$) \\
\hline 0.000 &  -0.443147\footnote{Exact solution $E_0=-\ln2+1/4$
given
by Bethe Ansatz.} & -0.443146 \\
0.001 & -0.443166 & -0.443165 \\
0.002 & -0.443196 & -0.443197 \\
0.005 & -0.443333 & -0.443333 \\
0.010 & -0.443655 & -0.443655 \\
0.020 & -0.444537 & -0.444537 \\
0.050 & -0.448374 & -0.448376 \\
0.100 & -0.457246 & -0.457246 \\
1/3   & -0.517954 & -0.517954 \\
\end{tabular}
\end{ruledtabular}
\end{table}
%%%%%%%%%%%%%%%%%%%%%%%%%%%%%%%%%%%%%%%%%%%%%%%%

%%%%%%%%%%%%%%%%%%%%%%%%%%%%%%%%%%%%%%%%%%%%%%%%%%%%%%%
\begin{figure}[t]
\includegraphics [width=0.45\textwidth]{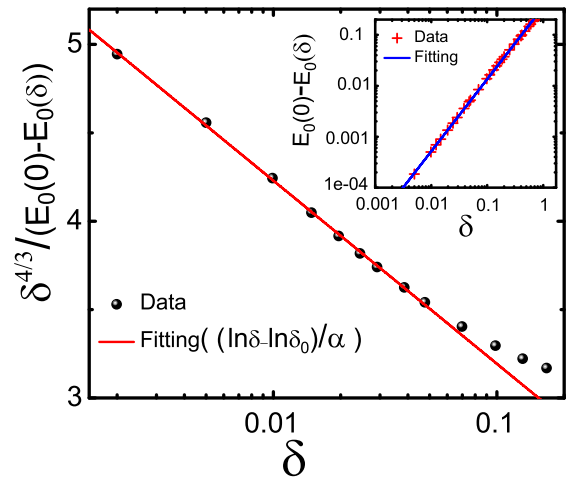}
\caption{ (Color online) Estimation of the constants $\alpha$ and
$\delta_0$ from iMPS data. In the inset, $E_0(0)-E_0(\delta)$ as a
function of $\delta$ is shown. } \label{fig3}
\end{figure}
%%%%%%%%%%%%%%%%%%%%%%%%%%%%%%%%%%%%%%%%%%%%%%%%%%%%%%%
 To understand more the critical behavior of the alternating Heisenberg
 chain in the view of how a bond-dimerized Heisenberg chain approaches
 the uniform limit $(J'=J)$ in our iMPS approach,
 we also calculate the groundstate by using the Hamiltonian
 $H= \tilde{J}\sum_{i} (1-\delta) \vec{S}_{2i-1}\cdot\vec{S}_{2i}+
 (1+\delta) \vec{S}_{2i}\cdot\vec{S}_{2i+1}$,
 written in terms of the dimerization parameter $\delta$.
 Here, $\tilde{J}=(J+J')/2$ and $\delta=(J'-J)/(J'+J)$ with
 $0\leq\delta\leq1$.
 In the Table~\ref{tab:energy}, a comparison between the groundstate
 energies $E_0$ from DMRG \cite{Kumar} and our iMPS approach is made
 for $\tilde J=(J+J')/2=1$ and the truncation dimension $\chi=56$.
 Note that our iMPS groundstate energies agree very well
 with those from DMRG.

 By virtues of a field theoretical approach,
 Black and Emery \cite{Black} have related the critical behavior to the four-state
 Potts model and found that the energy per site follows a power law with
 a logarithmic correction, i.e.,
 $E_0(0)-E_0(\delta)\simeq \alpha \delta^{4/3}/(\ln{\delta}-\ln\delta_0)$
 with the numerical fitting constants $\alpha \approx -2.2$ and $\delta_0 \approx 110$.
 By using DMRG, Papenbrock {\it et al.} \cite{Papenbrock} have also found
 that a power law ($\propto\delta^{1.45}$) is
 quite well in the range of $0.008\leq\delta\leq0.1$, while
 logarithmic correction is need for small $\delta$.
 From our iMPS representation, in Fig. \ref{fig3}, we plot
 the energy per site $E_0(\delta)$ with the dimerization parameter
 $\delta$ for $\tilde{J}=1$.
 $E_0(0)=1/4-\ln{2}$ is corresponding to the uniform limit.
 Figure \ref{fig3} shows
 the power law behavior with the logarithmic correction,
  $E_0(0)-E_0(\delta)\simeq \alpha \delta^{4/3}/(\ln{\delta}-\ln\delta_0)$
 with the numerical fitting constants $\alpha \approx -2.21$ and $\delta_0 \approx 114$.

\section{Topological quantum phase transition}
%%%%%%%%%%%%%%%%%%%%%%%%%%%fig 3%%%%%%%%%%%%%%%%%%%%%%%%%%%%%%%%
\begin{figure}
 \includegraphics[width=0.45\textwidth] {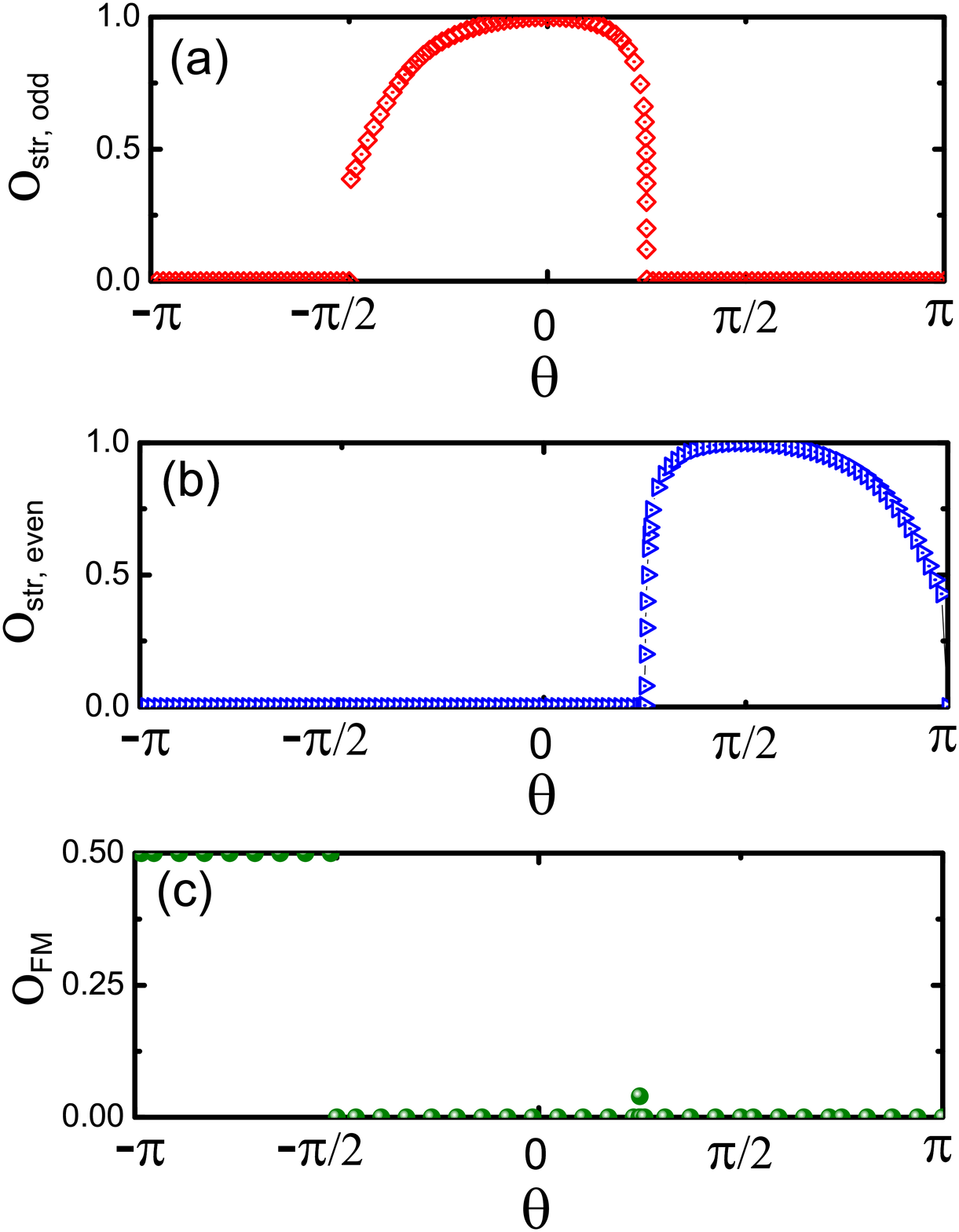}
 \caption{ (Color online)
 String order parameters $O_{str,odd/even}$ and ferromagnetic order
 parameter $O_{FM}$ as a function of $\theta$ with
 $J=\cos\theta$ and $J'=\sin\theta$.}
 \label{fig4}
\end{figure}
%%%%%%%%%%%%%%%%%%%%%%%%%%%%fig 3%%%%%%%%%%%%%%%%%%%%%%%%%%%%%%%%

\begin{figure}
\includegraphics [width=0.45\textwidth]{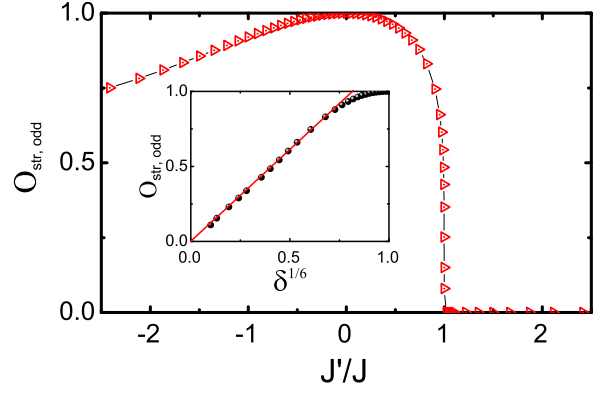}
\caption{ (Color online) String order parameter $O_{str,odd}$ as a
function of $J'/J$ with $J=1
$. In the inset, the string order
parameter $O_{str,odd}$ is plotted as a function of $\delta^{1/6}$.
The straight line, as a guide of eye, is also drawn.
} \label{fig5}
\end{figure}

 In the antiferromgantic Heisenberg Hamiltonian of Eq. (\ref{Hamt}) with the bond
 alternation, the transition between the two dimerized states at
 $J'/J=1$ does not involve any change of symmetry although the second derivative
 of the groundstate energy exhibits its singluar behavior indicating phase transition
 in Fig. \ref{fig3}.
 The absence of any explicit change of symmetry implies no local order
 parameters characterizing each phase.
 However, it manifests as a change of topological order, i.e.,
 string order.
 To see this,
 let us introduce two string order parameters based on the bond alternation:
\begin{subequations}
 \begin{eqnarray}
 %\nonumber
 O^\alpha_{str,even}  \!\!\! &=&  \!\!\!  \lim_{|i-j|\rightarrow\infty}
 \left(-4\langle
 S^\alpha_{2i} \exp [i\pi\sum_{k=2i+1}^{2j-2}S^\alpha_k] S^\alpha_{2j-1}\rangle \right), \\
 O^\alpha_{str,odd}  \!\!\!  &=&  \!\!\!
 \lim_{|i-j|\rightarrow\infty} \left(-4\langle S^\alpha_{2i+1} \exp
 [i\pi\sum_{k=2i+2}^{2j-1}S^\alpha_k] S^\alpha_{2j}\rangle \right),
 \label{equ:string}
\end{eqnarray}
\end{subequations}
where $\alpha=x$, $y$, and $z$.
 Due to the SU(2) rotationally invariant ground
 states, $O^x_{str}=O^y_{str}=O^z_{str}$.
 From our iMPS groundwavefunction, we directly calculate the defined string
 orders. In Fig. \ref{fig4}, the order parameters are
 displayed as a function of the angle variable $\theta=\tan^{-1}{J'/J}$.
 As we mentioned, the string orders are the saturated values of
 string correlations for a large lattice distance.
 Figure \ref{fig4} clearly shows that the odd string order
 is finite in the $ -\pi/2 < \theta < \pi/4$ phase and zero otherwise,
 while the even string order is finite in the $ \pi/4 < \theta < \pi$.
 These results imply that each system parameter range
 can be characterized by each long-range string order.
 Also, as the system parameter crosses the critical point $J'=J$,
 the system undergoes a TQPT
 between the two topologically ordered phases.
 Similar to the spin-1 Heisenberg chain understood by
 the hidden $Z_2\otimes Z_2$ breaking symmetry~\cite{Nijs,Oshikawa},
 a similar hidden symmetry breaking may occur for each phase.
 Further,
 we plot the string order parameter $O_{str,odd}$ as
 a function of $J'$ with $J=1$ in Fig. \ref{fig5}.
 In the inset of Fig. \ref{fig5},
 the string order parameter $O_{str,odd}$
 is plotted as a function of the dimerization parameter $\delta$.
 Note that the string order parameter
 $O_{str,odd}$ scales as $\delta^{1/6}$ ~\cite{Hida}.
 As a result, the critical exponent is given as
 $\beta = 1/12$ via $O_{str,odd} \propto \delta^{2\beta}$ ~\cite{Shelton}.
 Hence, the TQPT is in the same universality class with the Gaussian
 phase transition.

%%%%%%%%%%%%%%%%%%%%%%%%%%%%%%%%%%%%%%%%%%%%%%%%%%%%%%%
%%%%%%%%%%%%%%%%%%%%%%%%%%%%%%%%%%%%%%%%%%%%%%%%%%%%%%%

 When the $\theta$ approaches $-\pi/2$ or $\pi$,
 the system can be regarded as an effective
 spin-1 antiferromagentic Heisenberg chain
 because of the strong odd- or even-bond dimerizations.
 In these limits,
 the numerical values of our string orders $O_{str,odd/even}=0.3873$
 become very close to the results,
 for the antiferromagentic spin-1 Heisenberg chain,
 ${\cal O}^{\textrm{S=1}}_{str} = 0.37434447$
 from the iMPS approach \cite{Su} and
 ${\cal O}^{\textrm{S=1}}_{str}= 0.37432509$
 from the DMRG method \cite{Whitestring}.
 For $-\pi < \theta < -\pi/2$ ($ J' < 0$ and $J <0$),
 both the odd and even string orders
 become zero, while the ferromagnetic local order
 $O_{FM}=\langle S_{i}\rangle$ become finite, which indicates
 that the system is in the ferromagnetic phase.
 In the magnetization in Fig. \ref{fig4},
 one may notice a non-zero value of the local
 magnetization $\langle S_i \rangle$ at
 $\theta=\pi/4$, i.e., an isotropic Heisenberg chain that should have zero
 magnetization. The non-zero value of the magnetization
 at $\theta=\pi/4$ is due to finite truncation dimension $\chi$.
 However, it scales down to zero in the limit of infinite truncation
 dimension, i.e., the thermodynamic limit ~\cite{HongleiWang}.

\begin{figure}
\includegraphics [width=0.45\textwidth]{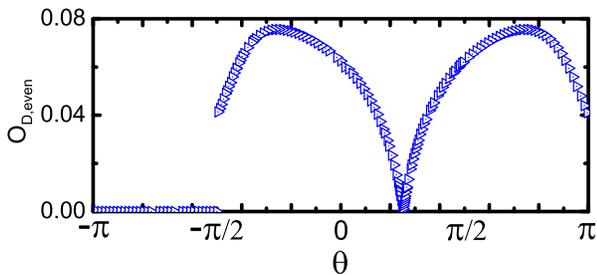}
\caption{ (Color online) Dimer orders $O_{D,even}$
 as a function of $\theta$ with
 $J=\cos\theta$ and $J'=\sin\theta$.
} \label{fig6}
\end{figure}

 Actually, Hida~\cite{Hida} introduced a similar string order parameter to argue
 a crossover between a Haldane-gap phase and a `dimer phase'
 in the spin-$1/2$ bond alternating
 Heisenberg chain for $J>J'$.
 However, his `dimer phase' is originated
 from the symmetry of the Hamiltonian rather than a broken symmetry of groundstate
 because both the Hamiltonian and its groundstate possess the same symmetry, i.e.,
 two-site translational invariance.
 In this sense, Hida has found the existence of a topological order for $J'>J$
 in the spin-$1/2$ bond alternating
 Heisenberg chain for $J>J'$.
 However, for $J>J'$, no crossover happens in the system because his `dimer phase'
 might be rather a `dimerized state' due to the bond alternation.
 To make this point clearer,
 we calculate the odd- and even- dimer orders
 defined by
\begin{subequations}
 \begin{eqnarray}
 %\nonumber
 O_{D, even}  \!\!\!  &=&  \!\!\!
 \lim_{|i-j|\rightarrow\infty} \langle O_{D}(2i)O_{D}(2j) \rangle,
 \\
 O_{D, odd}  \!\!\!  &=&  \!\!\!
 \lim_{|i-j|\rightarrow\infty} \langle O_{D}(2i+1)O_{D}(2j) \rangle,
 \label{equ:dimer}
\end{eqnarray}
\end{subequations}
 where
 $ O_{D}(i) = \vec S_{i} \cdot \vec S_{i+1} -\vec S_{i+1} \cdot \vec
 S_{i+2}$. In our system, we calculate the even- and odd- dimer orders
 $O_{D, odd}=-O_{D, even}$.
 In Fig. \ref{fig6}, the even-dimer order is displayed
 as a function of the angle variable $\theta$.
 Both the odd- and even-dimer orders are finite in the two Haldane
 phases, which implies that the two Haldane phases cannot be
 distinguished by the dimer orders
 although
 both the odd- and even-dimer orders disappear at the critical point $\theta = \pi/4$.
 These disappearances of the dimer orders at the critical
 point are originated from the symmetry of the Hamltionian
 because the Hamiltonian is one-site translational invariance
 and its groundstate has the same symmetry of the one-site
 translation invariance.
 Consequently, the dimer orders are not the order parameters
 characterizing the two Haldane phases
 although the dimer orders exist in the two Haldane
 phases.
 Hence, the two long-range string orders characterize
 the two Haldane phases, respectively, and the TQPT occurs between
 the two Haldane phases.

%%%%%%%%%%%%%%%%%%%%%%%%%%%%%%%%%%%%%%%%%%%%%%%%%%%%%%%
%%%%%%%%%%%%%%%%%%%%%%%%%%%%%%%%%%%%%%%%%%%%%%%%%%%%%%%

\section{von Neumann entropy and central  charge}
  At a quantum critical point,
  characteristic common singular behaviors
  of thermodynamics properties, depending only on few features
  such as dimensionality and symmetry, allow us to classify
  quantum phase transitions by using the concept of
  the universality classes, i.e.,
  a type of quantum phase transition.
  Especially, the central charge \cite{Cardy} can be used to classify
  a universality class of the TQPT at $J'/J=1$.
  In the iMPS representation, at a quantum critical point,
  one can obtain the central charge $c$
  via a so-called finite-entanglement scaling exponent $\kappa$
  numerically from the unique behaviors
  of the correlation length $\xi= a\chi^\kappa$
  and the von Neumann entropy $S=(c\kappa/6)\log_2{\chi}$
  \cite{Tagliacozzo}.
 Here, $a$ and $\chi$ are a numerical fitting constant and the
 truncation dimension, respectively.
 Thus, in order to get more insight on the TQPT, we will discuss
 the von Neumann entropy and the central charge at $J'/J=1$ in this section.

 Actually,
 the von Neumann entropy can quantify quantum entanglement as a good
 measure of bipartite entanglement between two subsystems of a pure
 state \cite{Osterloh,Amico}.
 In our iMPS representation,
 the spin chain can be partitioned into the two parts denoted by the
 left semi-infinite chain $L$ and the right semi-infinite chain $R$.
 The von Neumann entropy is defined as
 $S=-\mathrm{Tr}\varrho_L\log_2\varrho_L=-\mathrm{Tr}\varrho_R\log_2\varrho_R$
 in terms of the reduced density matrix $\varrho_L$ or $\rho_R$ of
 the subsystems $L$ and $R$. The von
 Neumann entropy for the semi-infinite chains $L$ or $R$ can be
 expressed as
\begin{equation}
 S_i=-\sum_{\alpha=1}^\chi \lambda_{i,\alpha}^2 \log_2 \lambda_{i,\alpha}^2,
 \label{entropy}
\end{equation}
 where $\lambda_{i,\alpha}$'s are diagonal elements of the matrix
 $\lambda$ that
 could be directly obtained in the iMPS representation.
 This is because, when one partitions the two semi-infinite
 chains $L(-\infty,\cdots,i)$ and $R(i+1, \cdots,\infty)$, one gets
 the Schmidt decomposition
 $|\Psi\rangle=\sum_{\alpha=1}^{\chi}\lambda_{i,\alpha} |\phi_L\rangle|\phi_R\rangle$.
 From the spectral decomposition, $\lambda_{i,\alpha}^2$ are actually
 eigenvalues of the reduced density matrices for the two
 semi-infinite chains $L$ and $R$. In our two-site translational
 invariant iMPS representation, there are two Schmidt coefficient
 matrices $\lambda_A$ and $\lambda_B$ that describe two possible ways
 of the partitions, i.e., one is on the odd sites, the other is on
 the even sites.
 From the $\lambda_A$ and $\lambda_B$,
 one can obtain the two von Neumann entropies depending on the
 odd- or even-site partitions.
%
%%%%%%%%%%%%%%%%%%%%%%%%%%%%%%%%%%%%%%%%%%%%%%%%%%%%%%%%%%%%%%%
\begin{figure}
  \includegraphics[width=0.45\textwidth]{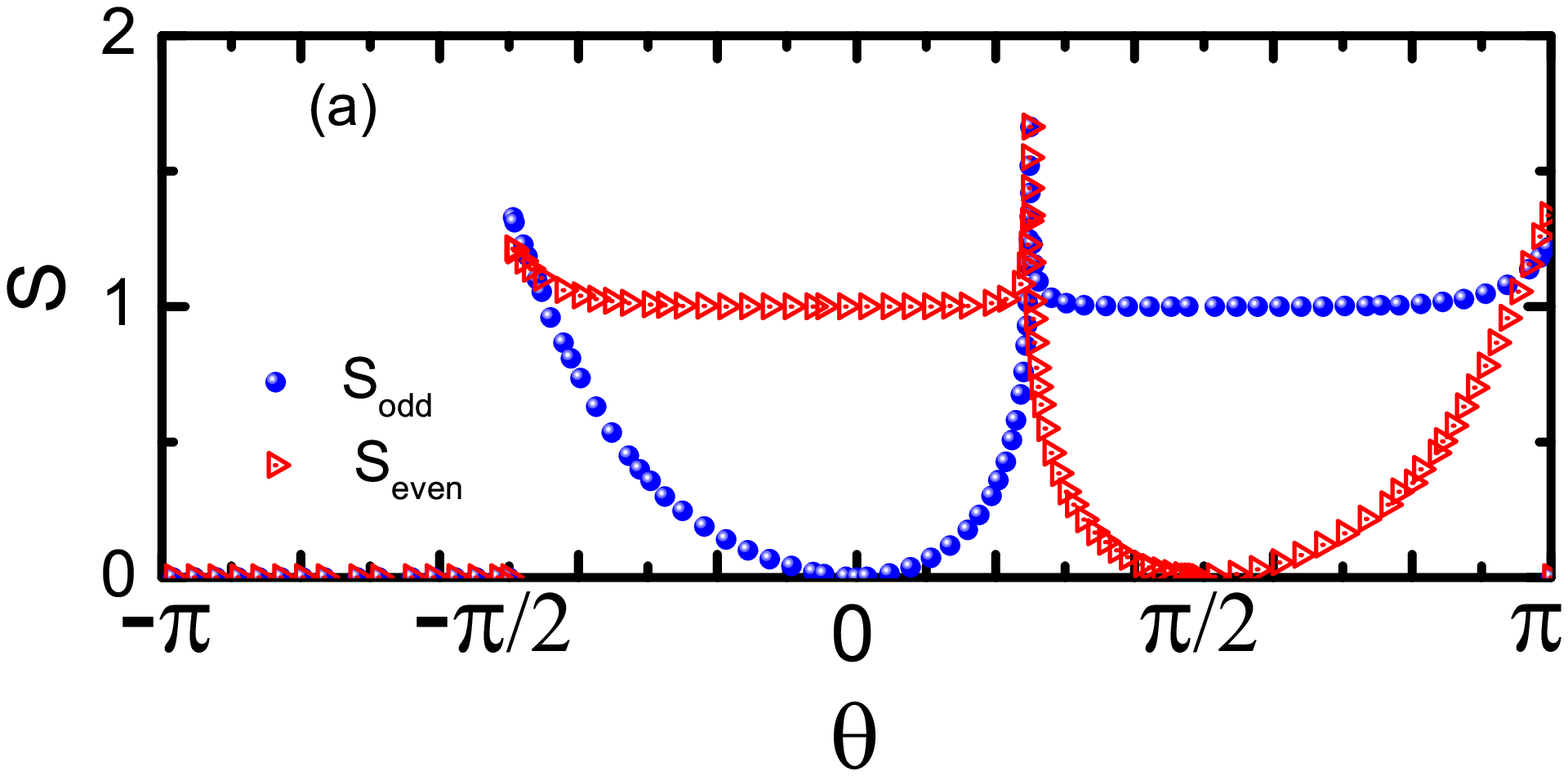}
  \includegraphics[width=0.45\textwidth]{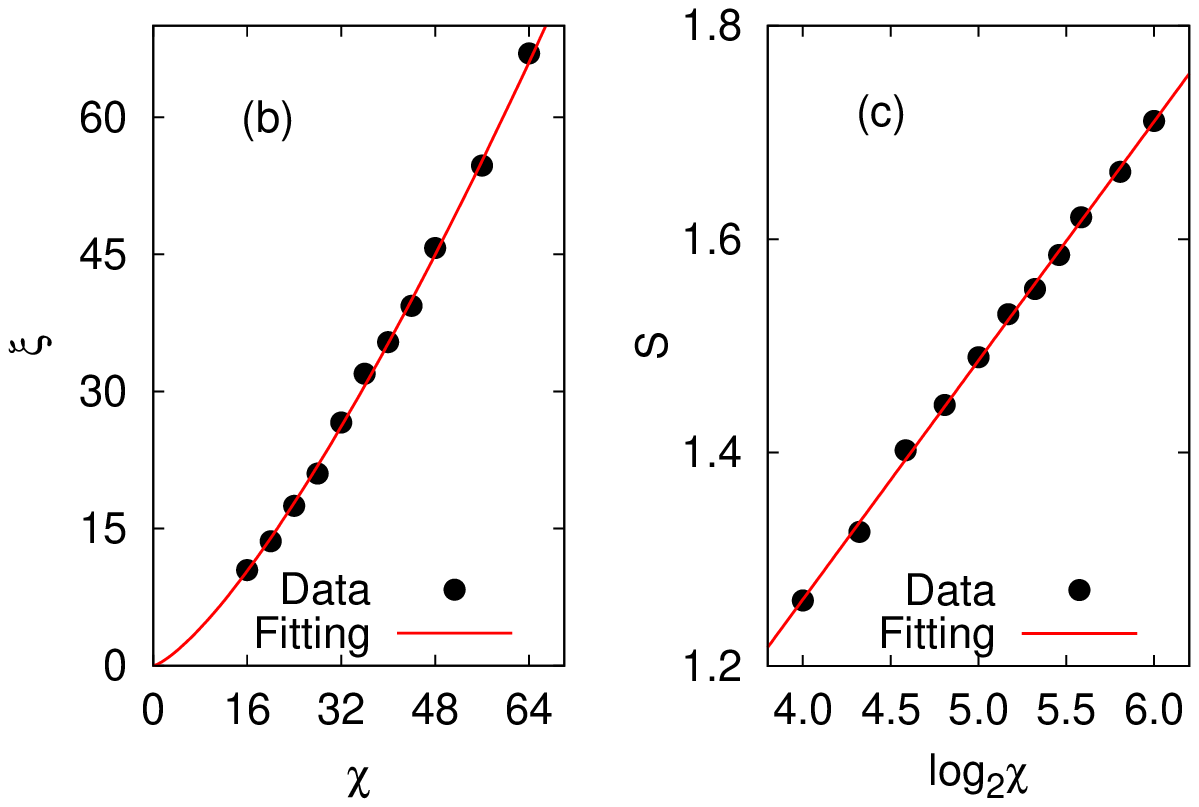}
 \caption{(Color online) (a) Von Neumann entropies $S_{even}$ and
 $S_{odd}$ as a function of the parameter $\theta$.
 The von Neumann entropies for odd- and even-bonds show a peak
  at $\theta=\pi/4$ and discontinue at $\theta = -\pi/2$, $\pi$.
  (b) Correlation length $\xi$ as a function of the
  truncation dimension $\chi$ at the critical point $\theta=\pi/4$.
   The power curve fitting $\xi=a \chi^ \kappa$ yields $a=0.255$ and $\kappa=1.336$.
  (c) Scaling of the von Neumann entropy $S$ with the truncation
  dimension $\chi$ at the critical point $\theta=\pi/4$, i.e., $J'=J$ corresponding
  to the conventional spin-$1/2$ AF isotropic Heisenberg chain.
  For $\kappa$ from (b), the linear fitting
  $S=(c\kappa/6)\log_2{\chi}+b$ yields $b=0.369$ and the central charge
  $c\approx 1.004$.
} \label{fig7}
\end{figure}
%%%%%%%%%%%%%%%%%%%%%%%%%%%%%%%%%%%%%%%%%%%%%%%%%%%

 In Fig.~\ref{fig7} (a), we plot the von Neumann entropies
 $S_{odd}$ and $S_{even}$
 as a function of the angle variable $\theta=\tan^{-1}J'/J$.
 Both the entropies for odd and even bonds show their singular behaviors at the
 same values $\theta =-\pi/2$, $\pi/4$, and $\pi$.
 Note that the singular behaviors of the von Neumann entropies correspond
 to the quantum phase transition points
 from the order parameters in Fig. \ref{fig4}.
 Hence, the von Neumann entropies
 give the same phase diagram from the order parameters.

 In Fig.~\ref{fig7} (b) and (c),
 the correlation length $\xi$ and the von Neumann entropy $S$ are plotted as a
 function of the truncation dimension $\chi$ at the critical point
 $\theta=\pi/4$ ($J'/J=1$).
 The truncation dimensions $\chi$ are taken from $16$ to $64$.
 It is shown that both the correlation length $\xi$ and the von Neumann
 entropy $S$ diverge as the truncation dimension $\chi$ increases.
 From a power-law fitting on the correlation length $\xi$, we have
 $\kappa = 1.336$ and $a = 0.255$. As shown in
 Fig.~\ref{fig7} (c), our numerical result demonstrates
 a linear scaling behavior, which gives a central charge $c \simeq
 1.004$ with $\kappa = 1.336$.
 Our central charge is close to the exact value $c = 1$
 predicted by conform field theory for an isotropic antiferromagnetic Heisenberg chain.
 Consequently, the TQPT between
 the odd- and the even- Haldane phases at $J'/J=1$
 is a Gaussian transition which is characterized by a central charge
 $c=1$ and the occurrence of a phase transition between two gapful
 phases. This result is consistent with the classification of the
 TQPT from the critical exponent $\beta=1/12$.

%%%%%
\section{Fidelity per lattice site}

%%%%%%%%%%%%%%%%%%%%%%%%%%%fig 1%%%%%%%%%%%%%%%%%%%%%%%%%%%%%%%%%%%%
\begin{figure}[t]
\includegraphics [width=0.45\textwidth]{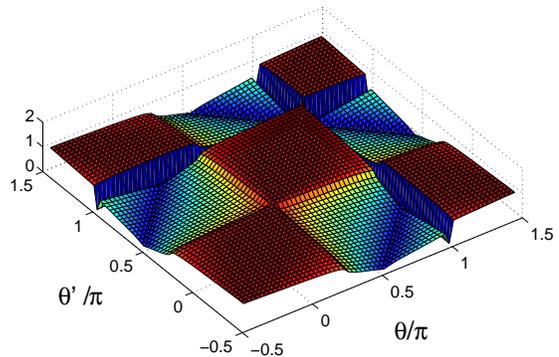}
\caption{ (Color online)
 Fidelity per site $d(\theta',\theta)$ surface as a
 function of two parameter $\theta'$ and $\theta$
 for a bond alternating Heisenberg chain with $J=\cos\theta$
 and $J'=\sin\theta$.}
 \label{fig8}
\end{figure}
%%%%%%%%%%%%%%%%%%%%%%%%%%%fig 1%%%%%%%%%%%%%%%%%%%%%%%%%%%%%%%%%%%%

 As we discussed in the previous section, the von Neumann entropy, as
 an entanglement measure, can detect the topological phase
 transition at $J'/J=1$. In this section, we study a fidelity per
 site as another universal indicator for quantum phase
 transition. From our iMPS groundstate wavefunction $| \Psi(\theta)\rangle$ in terms of
 the angle variable $\theta=\tan^{-1}J'/J$,
 one can define a fidelity $F(\theta',\theta)= |
 \langle\Psi(\theta')| \Psi(\theta)\rangle|$ between the groundstate
 wavefunctions for two different control parameters $\theta$ and $\theta'$.
 A fidelity per lattice site (FLS)\cite{Zhou} $d$  can be defined as
\begin{equation}
\ln d(\theta',\theta) = \lim_{N \rightarrow \infty} \frac {\ln
F(\theta',\theta)}{N}, \label{dinf}
\end{equation}
 where $N$ is the system size.
In Fig.~\ref{fig8}, the FLS $d(\theta',\theta)$ is plotted in the
two dimensional control parameter space. The three singular points
$(\theta', \theta)=(\pi/4,\pi/4)$, $(-\pi/2,-\pi/2)$ and $(\pi,\pi)$
are observed on the FLS surface, which indicates that there occur
quantum phase transitions when control parameters across these
values.
The continuous behavior of the FLS function across the ``pinch''
point $(\theta', \theta)=(\pi/4,\pi/4)$ implies that a continuous
quantum phase transition occurs for the topological phase
transition. The discontinuous behaviors of the FLS function at two
other points are corresponding to the first-order quantum phase
transitions~\cite{Zhou}.
The pinch points of the FLS
 correspond
 to the quantum phase transition points
 from the order parameters in Fig. \ref{fig4} and the von Neumann entropies in
 Fig. \ref{fig7}(a).
 The FLS
 gives the same phase diagram from the order parameters and the von Neumann entropy.
 Hence, it is shown that a TQPT can
 be detected by FLS.

\section{Conclusion}
 Quantum phase transitions have been investigated systematically in
 a spin-$1/2$ bond-alternating Heisenberg chain
 by using the iMPS representation with the iTEBD method.
 By calculating the odd- and even-string orders and
 the ferromagnetic order,
 the three phases, i.e., the odd- and even-Haldane phases
 and the ferromagnetic phases, were found in the plane of the bond-alternating
 interactions.
 The TQPT between the odd- and even-Haldane phases was
 classified as a Guassian-type phase transition from the central
 charge at the critical point and the critical exponent of the string orders.
 Also, it was clearly shown that the FLS and the von Neumann entropy
 can detect the TQPT.

\begin{acknowledgments}
 We thank Huan-Qiang Zhou and Jin-Hua Liu
 for helpful discussions. This work was supported
 by the Fundamental Research Funds for the Central Universities
 (Project No. CDJZR10100027) and the NSFC under Grant No.11104362.
\end{acknowledgments}

\end{document}